\documentclass[usenatbib,useAMS]{mn2e}
\usepackage{times}
\usepackage{epsfig}

\begin{document}
\title[A microlensing hunt for Earth-mass planets]{An anomaly detector with
immediate feedback to hunt for planets of Earth mass and
below by microlensing}
\author[M. Dominik et al.]{M. Dominik,$^{1}$\thanks{Royal Society University Research Fellow}\thanks{E-mail: md35@st-andrews.ac.uk}
N. J. Rattenbury,$^{2}$  A. Allan,$^{3}$
S. Mao,$^{2}$ D. M. Bramich,$^{4}$ \and M. J. Burgdorf,$^{5}$
E. Kerins,$^{2}$ 
Y. Tsapras,$^{5}$ 
and \L. Wyrzykowski$^{6,7}$ \\
$^{1}$SUPA, University of St Andrews, School of Physics \& Astronomy, 
North Haugh, St Andrews, KY16 9SS, United Kingdom\\
$^{2}$Jodrell Bank Observatory, Macclesfield, Cheshire, SK11 9DL, United Kingdom \\
$^{3}$School of Physics, University of Exeter, Stocker Road, Exeter
EX4 4QL, United Kingdom \\
$^{4}$Isaac Newton Group of Telescopes, Apartado de Correos 321, 38700
Santa Cruz de La Palma, Canary Islands, Spain \\
$^{5}$Astrophysics Research Institute, Liverpool John Moores University, Twelve Quays House, Egerton Wharf, Birkenhead, CH41 1LD, United Kingdom \\
$^{6}$Institute of Astronomy, University of Cambridge, Madingley Road, Cambridge CB3 0HA, United Kingdom \\
$^{7}$Warsaw University Astronomical Observatory, Al.~Ujazdowskie 4, 00-478
Warszawa, Poland}

\maketitle

\begin{abstract}
The discovery of OGLE~2005-BLG-390Lb, the first cool rocky/icy exoplanet, impressively demonstrated the
sensitivity of the microlensing technique to extra-solar planets 
below $10~M_\oplus$. A planet of $1~M_\oplus$ instead of the expected
$5~M_\oplus$ for OGLE~2005-BLG-390Lb (with an uncertainty factor of two)
in the same spot would have provided a detectable deviation with an
amplitude of $\sim 3$ per cent and a duration of $\sim 12~\mbox{h}$.
While a standard sampling interval of 1.5 to 2.5 hours for microlensing follow-up observations appears to
be insufficient for characterizing such light curve anomalies and 
thereby claiming the discovery of the planets that caused these,
an early detection of 
a deviation could trigger higher-cadence sampling which would have
allowed the discovery of an Earth-mass planet in this case. 
Here, we describe the implementation of an automated anomaly detector,
embedded into the eSTAR system,
that profits from immediate feedback provided by the robotic telescopes
that form the RoboNet-1.0 network.
It went into operation for the 2007 microlensing observing season.
As part of our discussion about an optimal strategy for planet detection,
we shed some new light
on whether concentrating on highly-magnified events is promising and
planets in the 'resonant' angular separation equal to the angular 
Einstein radius are revealed most easily. Given that sub-Neptune mass
planets can be considered being common around the host stars probed
by microlensing (preferentially M- and K-dwarfs), the higher number of events that can be monitored with a network of 2m telescopes and the 
increased detection efficiency for planets below $5~M_\oplus$ arising
from an optimized strategy gives a common effort of current microlensing
campaigns a fair chance to detect an Earth-mass planet (from the ground) ahead of the COROT or Kepler missions. The detection limit of gravitational microlensing extends even below $0.1~M_{\oplus}$,
but such planets are not very likely to be detected from current campaigns. However, these will 
be within the reach of high-cadence monitoring with a network of wide-field telescopes or a space-based telescope.
\end{abstract}


\section{Introduction}
After \citet{MP91} first pointed out that microlensing events can
be used to infer the presence of extra-solar planets or place limits
on their abundance, this technique has now become established with
several claimed detections \citep{Bond:planet,Udalski:planet,PLANET:planet,
Gould:planet}. The discovery of OGLE 2005-BLG-390Lb
\citep{PLANET:planet,Dominik:planet}, estimated to be 5 times more massive than Earth, with an uncertainty factor of two,
under the lead of the PLANET (Probing Lensing Anomalies NETwork)/RoboNet
campaign demonstrated that microlensing not only can detect massive gas giants, but also planets that harbour a rocky/icy surface under a thin atmosphere. Moreover, it provided the first 
observational hint that cool rocky/icy planets are actually quite
common, as previously predicted by simulations based on core-accretion models of planet formation \citep{IdaLin3}.

It was already estimated by \citet{BenRhie} that there is a 
non-negligible chance of 1--2 per cent for detecting an Earth-mass
planet located at about 2~AU from its host star by means of 
observing a few-per-cent deviation in a microlensing light curve. 
However, such a discovery requires 
photometric measurements on a few hundred microlensing events,
assuming that a fair fraction of the host stars are orbited
by such planets.

A sufficient number of events can only arise from monitoring dense fields of stars. With a probability of $\sim 10^{-6}$ for a star in the Galactic bulge being magnified by more than 34 per cent at any given time due to the bending of light caused by the gravitational field of an intervening foreground star \citep{KP94}, and such a microlensing event lasting of the order of a month, one namely needs to monitor $10^{7}$ to $10^{8}$ stars.
This was achieved by
microlensing surveys like OGLE (Optical Gravitational Lensing Experiment) \citep{OGLE:first}, MACHO (MAssive Compact Halo Objects) \citep{MACHO:first}, EROS (Exp\'{e}rience de la Recherche d'Objets Sombres)
\citep{EROS:first} and MOA (Microlensing Observations in Astrophysics) \citep{MOA:first} with a roughly daily sampling.
Moreover, all these surveys have been
equipped with real-time alert systems \citep{OGLE:alert,OGLE:alert2,
MACHO:alert,EROS:alert,MOA:alert} that notify the scientific community
about ongoing microlensing events. This allows to schedule 
follow-up observations that provide an increased photometric accuracy, 
a denser event sampling, and/or coverage during epochs outside the target visibility from the telescope site used by the respective survey campaign.

The PLANET (Probing Lensing Anomalies NETwork) 
collaboration\footnote{\tt http://planet.iap.fr} established
the first telescope network capable of round-the-clock nearly-continuous
high-precision monitoring of microlensing events \citep{PLANET:first}
with the goal to detect gas giant planets and to determine their
abundance. For being able to detect deviations of 5 per cent,
PLANET aims at a 1-2 per cent photometric accuracy. With a typical sampling interval of 1.5 to 2.5 hrs allowing a characterization of planetary
anomalies on the basis of at least 10-15 data points taken while
these last, the required exposure time then limits the number of events that can be monitored. For bright (giant) stars, exposure times of
a few minutes are sufficient, so that PLANET can monitor about 20 events each night or 75 events per observing season, but this reduces to about 6 events each night or 20 events per season for fainter stars, for which exposure times reach 20~min \citep{PLANET:EGS}. In 1999, MACHO and OGLE-II together provided about
100 microlensing alerts, out of which only 7 were on giant source stars.
This severely limited PLANET in its planet detection capabilities: 
rather than 75 events, only about 25 could be monitored per season. 
The OGLE-III upgrade, in effect from 2002, had a major impact on the
potential of microlensing planet searches, paving the way towards
the now nearly 1000 microlensing events per year 
provided by the alert systems of the OGLE\footnote{\tt http://ogle.astrouw.edu.pl/ogle3/ews/ews.html} and
MOA\footnote{\tt http://www.massey.ac.nz/\~{}iabond/alert/alert.html} surveys. The much larger number of events arising from 
this upgrade allowed OGLE itself to obtain meaningful constraints
on planets of Jupiter mass \citep{Tsapras:OGLElimits,Snodgrass:OGLElimits},
while OGLE and MOA have even demonstrated that such planets can in fact
be detected by their surveys \citep{Bond:planet}. However,
for studying less massive planets, their sampling is insufficient.
At the same time, the OGLE-III upgrade enabled PLANET to exploit its full theoretical capability, and moreover, it gave PLANET a reliable chance to detect planets of a few Earth masses provided that these are not rare around the stars that cause the microlensing events. 
The discovery of OGLE 2005-BLG-390Lb \citep{PLANET:planet,Dominik:planet} explicitly proved the sensitivity of the PLANET observations to planets
in that mass range.

Microlensing events are also regularly monitored by the MicroFUN 
(Microlensing Follow-Up Network) team\footnote{{\tt http://www.astronomy.ohio-state.edu/\~{}microfun/}}. However, rather than exploiting a permanent network, MicroFUN concentrates on particularly promising events
and activates target-of-opportunity observations should such an event be in progress. Besides 1m-class telescopes, their stand-by network includes a
larger number of small (down to 0.3m diameter) telescopes operated by amateur astronomers, which are well suited to observe
the peaks of events over which the source star makes a bright target.

Since the PLANET network is restricted in its capabilities of monitoring 
$\sim\,$25 per cent of the currently alerted events with the observational
requirements, the planet detection rate could be boosted by using
larger (2m) telescopes or clusters of 1m-class telescopes.
In fact, such an upgrade is required in order to obtain a sample that
allows a reliable test of models of the formation and evolution of
planets around K- and M-dwarfs.
RoboNet-1.0\footnote{\tt http://www.astro.livjm.ac.uk/RoboNet/}
\citep{RoboNet} marks the prototype of a network of
2m robotic telescopes, not only allowing a fast response time,
but also a flexible scheduling by means of the multi-agent contract
model provided by the eSTAR project\footnote{\tt http://www.estar.org.uk} \citep*{eStar,eStar1}.
eSTAR is a key player in the Heterogeneous Telescope Networks (HTN) consortium and
involved in the IVOA (International Virtual Observatory Alliance) standards process.

If one aims at the discovery of Earth-mass planets, the standard
follow-up sampling of 1.5~hrs usually does not produce
the amount of data required to characterize the 
corresponding signals, and with less frequent sampling one
even faces a significant risk of missing any hint for a deviation
from an ordinary microlensing light curve. 
However, planets of Earth mass and even below can be discovered
by shortening the sampling interval to $\sim$\,10~min once a regularly sampled point is suspected to depart from a model light curve that represents a system without planet.
In order to properly trigger such anomaly alerts, all incoming data 
need to be checked immediately, and prompt action needs to be taken 
within less than $\sim$\,15~min. The amount of data and
the required response time for achieving a good 
detection efficiency for Earth-mass planets are however prohibitive
for relying on human inspection. Therefore,
we here describe the implementation of an automated anomaly detector that exploits the opportunities of immediate response and flexible scheduling of a network of robotic telescopes. A first similar warning system, dubbed EEWS, had been installed by OGLE in 2003 \citep{OGLE:alert2}, which
however involves further human inspection and operates with a 
single telescope. In contrast, our design
needs to succeed without any human intervention and take care of a heterogeneous telescope network. The underlying algorithm follows
previous experience on the assessment of anomalies. We explicitly aim at reaching a significant detection efficiency to
Earth-mass planets with the current survey/follow-up strategy of 
microlensing planet searches.

This paper is organized as follows. In Sect.~\ref{sec:ordinary} we
describe the modelling of ordinary microlensing events with particular
emphasis on the importance of robust parameter estimates, not confused
by outliers, in order to properly identify real deviations. While
Sect.~\ref{sec:lowmass} deals with the general strategy for detecting
low-mass planets by microlensing, we derive a suitable concept
for an anomaly detector in Sect.~\ref{sec:concept}. 
The embedding of the {\sc SIGNALMEN} anomaly detector, that went into
operation for the 2007 microlensing campaign, into the eSTAR project 
is discussed in Sect.~\ref{sec:embed}, before its algorithm is described 
in Sect.~\ref{sec:algo}. Sect.~\ref{sec:prospects} then discusses the prospects
of the {\sc SIGNALMEN} anomaly detector for discovering planets
of Earth mass and below. In Sect.~\ref{sec:summary}, we provide
a short summary and final conclusions.

The Appendix makes a point on the
inability to detect planets at the resonant separation in some
of the observed events that was not discussed earlier.

\section{Ordinary light curves and anomalies}
\label{sec:ordinary}
The bending of light due to the gravitational field of a 
foreground '{\em lens}' star with mass $M$ at distance $D_\mathrm{L}$ causes an observed background '{\em source}' star at distance
$D_\mathrm{S}$ to be magnified by \citep{Ein36}
\begin{equation}
A(u) = \frac{u^2+2}{u\,\sqrt{u^2+4}}\,,
\end{equation}
if both objects are separated on the sky by the angle 
$u\,\theta_\mathrm{E}$ with $\theta_\mathrm{E}$ denoting the
{\em angular Einstein radius}
\begin{equation}
\theta_\mathrm{E} = \sqrt{\frac{4GM}{c^2}\,(D_\mathrm{L}^{-1} -
D_\mathrm{S}^{-1})}\,.
\end{equation}

With the assumption that lens and source star move uniformly, where
$\mu$ is the absolute value of their relative proper motion,
the separation angle can be parametrized as
\begin{equation}
u(t) = \sqrt{u_0^2 + \left(\frac{t-t_0}{t_\mathrm{E}}\right)^2}\,,
\end{equation}
with $u_0$ denotes the closest approach at epoch $t_0$, and
$t_\mathrm{E} = \theta_\mathrm{E}/\mu$ is a characteristic
event time-scale.

Each set of observations with a specific telescope
and filter comprises a data archive $s$ of observed fluxes $F_i^{[s]}$ and their error bars $\sigma_{{F_i}^{[s]}}$ at epochs $t_i^{[s]}$.
With the source flux $F_\mathrm{S}^{[s]}$
and background flux $F_\mathrm{B}^{[s]}$ depending on the data
archive $s$, one observes symmetric light curves
\begin{equation}
F^{[s]}(t) = F_\mathrm{S}^{[s]}\,A[u(t)] + F_\mathrm{B}^{[s]}
\label{eq:lc}
\end{equation}
peaking at $t_0$.

Estimates for $(t_0,t_\rmn{E},u_0,F_\rmn{S}^{[s]},F_\rmn{B}^{[s]})$
can then be obtained by minimizing
\begin{equation}
\chi^2 = \sum_{k=1}^{m} \sum_{i=1}^{n^{[k]}}
\left(\frac{F^{[k]}(t)-F_i^{[k]}}{\sigma_{F_i^{[k]}}}\right)^2\,.
\end{equation}
While we use the CERN library routine {\tt MINUIT} for determining
$(t_0, t_\rmn{E}, u_0)$,
the source and
background fluxes $F_\mathrm{S}^{[s]}$ and $F_\mathrm{B}^{[s]}$ 
for any choice of $(t_0, t_\rmn{E}, u_0)$ simply
follow from linear regression as
\begin{eqnarray}
F_\mathrm{S} & = &\frac{\sum\frac{A(t_i)
 F_i}{\sigma_i^2} \sum\frac{1}{\sigma_i^2} -
\sum\frac{A(t_i)}{\sigma_i^2} \sum\frac{F_i}{\sigma_i^2}}
{\sum \frac{[A(t_i)]^2}{\sigma_i^2} \sum\frac{1}{\sigma_i^2}
- \left(\sum\frac{A(t_i)}{\sigma_i^2}\right)^2}\,, \nonumber \\
F_\mathrm{B} & = &\frac{\sum\frac{[A(t_i)]^2}
 {\sigma_i^2} \sum\frac{F_i}{\sigma_i^2} -
\sum\frac{A(t_i)}{\sigma_i^2} \sum\frac{A(t_i) F_i}{\sigma_i^2}}
{\sum \frac{[A(t_i)]^2}{\sigma_i^2} \sum\frac{1}{\sigma_i^2}
- \left(\sum\frac{A(t_i)}{\sigma_i^2}\right)^2} \,,
\end{eqnarray}
where the summations
 run from 1 to $n^{[k]}$, $\sigma_i \equiv \sigma_{F_i}$,
and the index $[k]$ has been dropped. Any archive $s$ can only be
included if it contains at least 3 data points.

The characteristic form of the light curve described by Eq.~(\ref{eq:lc})
is based on the assumption that both source and lens star are single
point-like objects that are moving uniformly with respect to each other as seen from Earth. Apart from planets orbiting the lens star, 
significant deviations, so-called {\em anomalies},
can however also be caused by binarity or multiplicity of lens or source, the finite angular size of the stars, or the revolution of the Earth (parallax effect).

Since it is our primary goal to detect light curve anomalies, it is essential to ensure that our adopted model is reasonably correct.
However, frequently our data do not allow strong constraints
to be placed on the model, in particular during early phases
of the event.
It is a well-known fact that OGLE announce a fair fraction of their
events with the prediction of quite high peak magnification,
whereas it turns out later that most of these peak at much lower magnifications. As studied in some detail by \citet{Albrow:Esti},
this is related to the fact that $\chi^2$-minimization is equivalent to obtaining a maximum-likelihood estimate of the model parameters if the
data are assumed to follow a Gaussian distribution, which
is biased, i.e.\ its expectation value does not coincide with the true expectation value of the considered quantity. Using the statistics of 
previously observed OGLE events, a Bayesian estimate that can
be obtained by adding an effective penalty function to $\chi^2$
comes closer to the expectation value \citep{Albrow:Esti}.
While the estimated value can be tuned by this, one does not fully get around the problem of large indeterminacy of the model parameters. 

A further problem arises from the necessity to avoid that our model
is driven towards data outliers. Otherwise, real anomalies would
be missed while points matching an ordinary light curve would seem
deviant. As a consequence, we would face the problem of not being able
to distinguish between ongoing anomalies and further data requiring
an adjustment of model parameters. Therefore, we apply a more sophisticated
algorithm for estimating the model parameters that is rather
invulnerable to outliers.

The model can be made to follow the bulk of the data by 
downweighting points according to their respective residual \citep*[e.g.][]{robustbook} as follows.
With the residuals
\begin{equation}
r_i^{[k]} = \frac{F^{[k]}(t)-F_i^{[k]}}{\sigma_{F_i^{[k]}}}
\end{equation}
and the median of their absolute values ${\tilde r}^{[k]}$
for each data archive, we give further (bi-square) weight
\begin{equation}
w_i^{[k]} = \left\{\begin{array}{ccl}
\left[1-\left(\frac{r_i^{[k]}}{K\,{\tilde r}^{[k]}}\right)^2\right]^2
& \mbox{for} & |r_i^{[k]}| < K\,{\tilde r}^{[k]} \\
0 & \mbox{for} & |r_i^{[k]}| \geq K\,{\tilde r}^{[k]}
\end{array}\right.
\label{eq:weight}
\end{equation}
to each data point, where we adopt $K = 6$ for the tuning constant. 
The choice of the weights, 
Eq.~(\ref{eq:weight}), means
that data points whose absolute residuals exceeds $K$ times their
median are ignored.
This procedure is repeated until the formal $\chi^2$
converges. However, we need to deal with non-linear models which are
prone to several possible $\chi^2$ minima. In contrast to linear models,
it can therefore happen that this procedure leads to periodic 
switching between different minima, where nevertheless a subsequence
converges to each of these. In this case, we have to live with the
absence of a unique minimum and choose that one with the lowest
$\chi^2$. With the formal $\chi^2$ not being dominated by outliers,
we can also reliably adjust the relative weight between different
data archives $k$ after each iteration step,
so that all $(\chi^2)^{[k]}/n^{[k]}$ coincide, preventing the
estimation of model parameters being influenced by the collective
over- or underestimation of error bars.

\section{Detection of low-mass planets}
\label{sec:lowmass}
It was pointed out by \citet{MP91} that planets orbiting the lens star
can reveal their existence by causing significant deviations to 
microlensing light curves. They also found that the probability to 
detect a planet becomes resonant if the angular separation from its
host star is comparable to the angular Einstein radius $\theta_\rmn{E}$, which reflects the fact that the detection of planets is aided by the
tidal field of their host star.
However, as pointed out in the Appendix, for a given event, in
particular for larger impact parameters, the detection probability of
smaller planets can actually drop to zero for angular separations
close to $\theta_\rmn{E}$ rather than reaching a maximum. In such case,
only slightly wider or closer separations can be probed.
It is a lucky coincidence that the gravitational radius of stars and distances within the Milky Way combine in such a way that the
angular Einstein radius converts to a projected separation $D_\rmn{L}\,\theta_\rmn{E} \sim 2~\mbox{AU}$ for $M = 0.3~M_\odot$, the typical mass of the lens stars, assuming $D_\rmn{S} \sim 8.5~\mbox{kpc}$ and
$D_\rmn{L} \sim 6.5~\mbox{kpc}$. 
\citet{GL92} quantified the prospects for detecting planets from
microlensing signatures by finding that Jupiter-mass planets distributed uniformly within angular separations
$0.6~\theta_\rmn{E} \leq d\,\theta_\rmn{E} \leq 1.6~\theta_\rmn{E}$, comprising the so-called {\em lensing zone}, have a probability of
15 per cent of being detected among microlensing events with peak
magnifications $A_0 \geq 1.34$, corresponding to the source entering
the Einstein ring (of angular radius $\theta_\rmn{E}$) of the lens star,
i.e.~$u_0 \leq 1$. As shown by \citet{GS:HME}, this probability increases
significantly if one restricts the attention to events with larger
peak magnifications, where about 80 per cent is reached for $A_0 \geq 10$.
Since the area subtended on the sky by angular source positions that 
correspond to a significant deviation decreases towards smaller
planet masses, both a shorter duration of the planetary signal and a 
smaller probability to observe it result. In contrast, the 
signal amplitude is only limited by the finite angular size of the source,
where significant signal reductions start arising once it becomes
comparable or larger than the size of the region for which a point source 
provides a significant deviation. However, \citet{BenRhie} estimated that Earth-mass planets still have a 1--2 per cent chance of providing a signal in excess of a few per cent.

Planets around the lens star affect the light curve only by means
of two dimensionless parameters, namely the
planet-to-star mass ratio $q$ and the separation parameter $d$, where
$d\,\theta_\rmn{E}$ is the instantaneous angular separation of the
planets from its host star (i.e.\ the lens star). With typical relative
proper motions between lens and source stars of
$\mu \sim 15~\umu\mbox{as}\,\mbox{d}^{-1}$,
microlensing events on Galactic bulge stars are usually 
observable for about a month or two, whereas planetary deviations last between
a few hours and a few days, depending on the mass of the planet. In contrast
to other indirect techniques, microlensing therefore obtains a snapshot measurement
of the planet rather than having to wait for it to complete its orbit. 
This gives microlensing the unique capability of probing
planets in wide orbits whose periods otherwise easily exceed the life-time of a project or its investigator.

With many events on offer from the OGLE and MOA surveys and only limited
resources available for follow-up observations, one needs to make a
choice which of these to monitor and how frequently to sample each event.
With the goal to maximize the number of detections of planetary deviations,
a prioritization algorithm that spreads the available observing time
over the potential targets has been devised by \citet{Horne:priority},
which forms a central engine of the RoboNet observing strategy.
Any such strategy must be based on observables, model parameters arising
from the collected data, or any other data statistics. 
As \citet{Horne:priority} pointed out, each data point
carries a detection zone with it, composed of the angular positions for
which a planet would have caused a detectable deviation.
Unless finite-source effects begin diminishing the detectability of
planets \citep{Han:priority}, detection zones grow with the current
magnification. Moreover, the same photometric accuracy can be 
achieved with smaller exposure times for brighter targets. 
An efficient prioritization algorithm therefore
needs to be based on both the current magnification and brightness along 
with the time when the last observation was carried out, where
taking into account the latter avoids obtaining redundant information.
Such a prioritization of events however does not consider how well
an observed deviation allows to constrain its nature of origin
and it also assumes that the model parameters of the ordinary
light curve are known exactly.

If the effect on the microlensing light curve is dominated by a single
planet, the lens system can be fairly approximated as a binary system
consisting of the star and this planet. Gravitational lensing by a
binary point-mass lens has been studied in great detail for equal
masses by \citet{SchneiWei:twomass} and later generalized for arbitrary
mass ratios by \citet{Erdl}. On the other hand, \citet{CR} have
discussed lensing by bodies of different mass scales. While their
target of interest was the brightness variation of 
individual images of QSOs that are gravitationally lensed by
an intervening galaxy, a very
similar situation arises for planets orbiting a lens star.
Similarly to individual stars in the galaxy splitting an image due
to lensing by the galaxy as a whole into 'micro-lensing', a planet
can further split one of the two images due to lensing by its host star
if it roughly coincides in angular position with that image.
\citet{Do99:CR} has further investigated the transition towards 
extreme mass ratios and shown how the case described by \citet{CR},
the so-called {\em Chang-Refsdal} lens, is approached. The derived
expansions into series have later been used by \citet{Bozza} for
discussing the case of multiple planets. Binary lenses in general
and planetary systems in particular create a system of extended
caustics, consisting of the angular positions for which a point-like
source star would be infinitely magnified. While sufficiently small
sources passing the caustics can provide quite spectacular signals, planets
are more likely to already reveal their existence on entering a much
larger region surrounding these.

For less massive planets, there are usually two separate regions for positions of the source star that lead to detectable planetary signals,
which are related to two types of caustics. Only if the angular separation of the planet from its host star is in a close vicinity to the angular Einstein radius $\theta_\rmn{E}$, where the corresponding range is
broader for more massive planets, a single caustic results and these
regions merge. Otherwise, there are 
one or two {\em planetary caustics} which are located around positions
for which bending of its light due to the gravitational field of the
lens star causes the source to have image at the position of the planet,
and a {\em central caustic} which can be found near the 
lens star \citep{GS:HME,Do99:CR}.
As \citet{Bozza} demonstrated, the planetary caustics associated
with different planets are almost always separated and any kind of interference between these is quite unlikely.
In contrast, \citet{GNS} pointed out that the central caustic
is always affected by the combined action of all planets.
However, it is likely, although not guaranteed, that there is a
hierarchical order among the effects of different planets, so that a
linear superposition is a fair approximation
\citep{Ratt:high,Han:central}.

While the absence of any deviations near the peak of extreme highly-magnified ordinary events that are related to the source potentially approaching the central caustic poses strict limits on the abundance of low-mass planets \citep{Abe:bestlimits,Dong:ray}, their actual discovery from this kind of deviations suffers from several complications. 
While the linear size of
the detection region around planetary caustics scales with the square root of the planet mass, it is proportional to the planet mass itself for the central caustic \citep{CR,GS:HME,Do99:CR,Chung,Han:caustics}. Therefore, the finite angular size of
the source star is more likely to cause a significant reduction of the signal amplitude.
Moreover, the characterization of the nature of origin for such deviations is significantly more difficult than for deviations related to 
planetary caustics. The latter provide further information by means of the time elapsed between the peak of the background ordinary light curve and the deviation, whereas central-caustic deviations involve a higher
degree of model degeneracies with more prominent finite-source and
parallax effects.
In any case, a promising sensitivity to Earth-mass planets is only reached
for lens-source impact parameters $u_0 \la 5\times 10^{-4}$, which
occur at a rate of less than one per year.

\begin{figure}
\includegraphics[width=84mm]{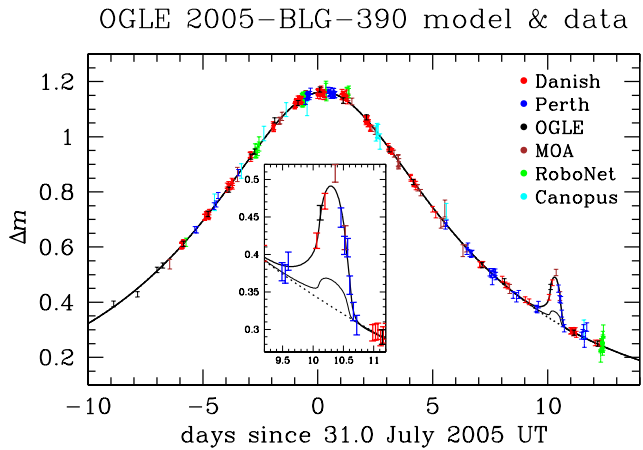}
\caption{Model light curve of microlensing event OGLE~2005-BLG-390 along with data
taken with the Danish 1.54m at ESO LaSilla (Chile), red, the Perth 0.6m (Western Australia), blue,
and the Canopus 1.0m (Tasmania), cyan, by PLANET, the Faulkes North 2.0m (Hawaii), green, by RoboNet-1.0, the
OGLE 1.3m (Chile), black, and the MOA 0.6m (New Zealand), brown,
where $\Delta m = 2.5\,\lg A(t)$ has been plotted along with
$m_i = 2.5 \lg A_i$. The $\sim\,15$ per cent deviation
lasting about a day revealed the existence of a 
planet with $m \sim 5.5~M_{\oplus}$ (uncertain to a factor two), while an Earth-mass planet in the same spot would have caused a 3 per cent deviation lasting about 12 hours (thin line). The time-scale of this event is $t_\mathrm{E} = 11.0~\mbox{d}$, while
$d = 1.610$ and $q = 7.6\times10^{-5}$. Moreover, $u_0 = 0.359$,
$t_0 = 31.231~\mbox{July}~\mbox{2005}~\mbox{UT}$, and the 
angle between the vector from the planet to its host star and the
source trajectory is $\alpha = 157.9\degr$, where the less centre of mass is
to the right hand side. Finally, the source star moves by
its own radius relative to the lens within $t_\star = 0.282~\mbox{d}$.
The dotted line refers to 
a model light curve in the absence of a planet.}
\label{fig:OB05390}
\end{figure}

On the other hand, the non-negligible probability to detect planetary signals if the source passes in the vicinity of planetary caustics 
offers a fair chance of detecting a planet of Earth-mass by also making
use of the large number of events that exhibit lower magnifications at a given time. Given these facts, it is not a surprise that the first sub-Neptune mass planet whose existence could be reported
on the basis of microlensing observations, OGLE~2005-BLG-390Lb \citep{PLANET:planet}, produced a 15 to 20 per cent signal at a magnification $A \sim 1.3$ about 10 days after 
an observed peak at magnification $A_0 \sim 3$ 
(see Fig.~\ref{fig:OB05390}) rather than a deviation within a
highly-magnified peak.

While the mass of OGLE~2005-BLG-390Lb is about $5~M_\oplus$, uncertain
to about a factor of two \citep{Do:Estimate2}, a planet of $1~M_\oplus$ in
the same spot would still have produced a signal with an amplitude 
of $\sim 3$ per cent, lasting $\sim 12~\mbox{h}$ rather than about
twice that long. The actual sampling would have been insufficient
for discovering such a planet in this configuration, but the situation 
would have been different had we decreased our sampling interval
to 10-15~min on the suspicion of a first deviation. This case 
explicitly shows how an anomaly detector can help us in not missing 
short-lasting small deviations (related to low-mass planets). By
requiring an initial sampling that is just dense enough for 
an ongoing anomaly being alerted before most of it has passed,
it moreover allows to monitor a sufficient number of events for
providing a reasonable number of planet discoveries.
The main gain of the anomaly detector will indeed be achieved
for detecting planets from perturbations related to planetary caustics at lower and moderate magnification, whereas a high-cadence
sampling can already be scheduled a-priori for (predictable) high magnifications without the need for any further alert.

The ability of detecting an anomaly depends on how well earlier data
constrain the model describing an ordinary light curve.
For large model parameter uncertainties, it 
becomes hard to distinguish a real deviation from a necessary 
model revision due to a previous misestimate,
for which $\chi^2$ adjustments are not a reliable indicator due
to the intricate parameter space and poor knowledge about the 
measurement uncertainties. Therefore, the anomaly detection
is more efficient after the peak of a microlensing has passed rather
than prior to it \citep*[c.f.][]{OGLE:alert2}, where the ability
is particularly vulnerable to data gaps.
Thus, if the increased detection efficiency for low-mass planets
that is achieved by means of the anomaly detector is a relevant
goal for a monitoring strategy, it is sensible to give preference
to events past peak over those pre peak for comparable
magnifications.
Although it is more difficult to decide whether a deviation from a
previous model is real or due to a model misestimate if constraints
on its parameters are weaker, it is more likely that a suspected
deviation occurs and is reported. This has the by-effect that more data
will be collected in this case, which in turn strengthens the 
model parameter constraints.
Despite the fact that the higher magnification around the peak allows
for accurate data being taken with shorter exposure times, the weak constraints on the position of the peak make it rather difficult to
detect an ongoing anomaly there, unless the peak region is
monitored quite densely and no data gaps occur.

\section{Concept for an anomaly detector}
\label{sec:concept}

If reported data deviate from the expected light curve, this 
could either mean that there is a real effect, the deviation
could be of statistical nature, or the data could
simply be erratic by any means. It is therefore impossible to arrive at
an appropriate judgement about the presence of anomalies on the basis of
a single deviating data point. However, such a point
should raise suspicion that an anomaly is indeed ongoing. Our anomaly detector, dubbed {\sc SIGNALMEN}, profits from the fact that real-time photometry and robotic telescope operation allow immediate feedback.
Rather than having to rely on a fixed sampling rate for a given event,
we can request prompt further observations once the modelling of
incoming data indicates a deviation from an ordinary light curve.

Based on the collected data, the anomaly detector can arrive at one
out of three possible conclusions and assign a corresponding status to the
event:
\begin{itemize}
\item there is no ongoing anomaly (ordinary)
\item there is an ongoing anomaly (anomaly)
\item not sure what is going on (check)
\end{itemize}
While the last option, corresponding to a suspected, unconfirmed
anomaly, does not look appealing at first sight, it actually marks
the strength of the feedback concept. In this case, we urgently 
request further observations on the same target, thereby providing
the anomaly detector with further data on which it can base the
decision in subsequent runs. In a 'recheck and repeat' strategy,
data whose absolute model residual is among the largest trigger further observations, and this process is repeated until a
decision about whether there is an anomaly can be taken with the desired significance.

The art of optimizing an anomaly detector is in finding the appropriate
balance between not missing planetary anomalies and avoiding false alerts. The availability of immediate feedback opens the possibility of using a rather low initial trigger level on the first suspicion of an anomaly,
which gives us a fair chance of detecting low-amplitude anomalies at
an early stage. The early detection is a vital feature for being able
to discover Earth-mass planets. In contrast, we do not care that much
about the detection of anomalies that have already been missed or are mostly over. 
A low initial trigger however means that we will need to spend 
a significant amount of time on collecting
evidence against the presence of an anomaly if the point that triggered
the 'check' observations does not constitute a real deviation. 
As pointed out in more detail in the following section, we aim at rechecking 5 per cent of the incoming
data for anomalous behaviour, while about 4 to 5 further points are expected to be required for providing sufficient evidence against.
This means
that we spend about 20 per cent of our observing time on checking 
potential anomalies. By basing the criterion for a significant 
deviation on a comparison of the model residual of the tested
data point with those of earlier data, we pay respect to the fact
that the true scatter of data is not properly reflected by the size of the reported error bars and can be non-Gaussian.

We also account for the fact that data collected with different telescopes may arrive in blocks rather than point-by-point and not necessarily
in time sequence. Moreover, all data are subject to change, which
not only means that reported $(F_i,\sigma_{F_i})$ might alter between
two runs of the anomaly detector, but data at certain epochs might
disappear, whereas additional data at other epochs prior to the
most recent data point might be released. 
By not making any distinction between whether 'new' data are released
in a block or arise from recent point-by-point observations, we
also take care of the possibility that an anomaly is already apparent in
the latest data update.

Our robust fitting scheme is rather powerful in identifying outliers and
therefore gives us some protection against failures of the real-time
photometry and weird results that might be the consequence. We have
implemented a further test for distinguishing between havoc photometry
and ongoing anomalies which produces an alert urging to check the data reduction. However, there is no way getting around the point that the capabilities of the anomaly detector will rise or fall with the quality of the real-time data analysis. In principle, one can also investigate correlations with observing conditions such as the reported seeing or sky brightness. However, such information may not be provided for all considered sites, so that we try to avoid relying on it as long
as possible. 

\section{Anomaly detector embedding and external interfaces}
\label{sec:embed}

The intelligent-agent architecture of the eSTAR project
constitutes the harness inside which the {\sc SIGNALMEN} anomaly detector operates. Thereby, it provides autonomous decision-making by means of software, which allows to build systems that learn and adapt. The eSTAR system 
provides the feedback loop by feeding the {\sc SIGNALMEN} anomaly 
detector with real-time data, which then replies with an expert opinion
that allows the eSTAR system to solve the distributed-scheduling problem of how to distribute follow-up requests over the network in order to
maximize the chances of detecting and characterizing an extra-solar planet.

The eSTAR project serves as a meta-network between existing proprietary robotic telescope networks
built upon a peer-to-peer agent based architecture \citep{Wooldridge}, which cuts across traditional notions that running such a network requires a ``master scheduler''. Instead, eSTAR can be viewed as a collaborative multi-agent system using a contract model. The crucial architectural distinction of such a system is that both the software controlling the science programme and those embedded at the telescope acting as a high-level interface to the native telescope control software are equally seen as ``agents''. A negotiation takes place between these agents in which each of the telescopes bids to carry out the work, with the user's agent scheduling the work with the agent embedded at the telescope that promises to return the best result. 
This preserves the autonomy of individual telescope operators to implement scheduling of observations at their facility as they see fit, and offers adaptability in the face of asynchronously arriving data. For instance, an agent working autonomously of the user can change, reschedule, or cancel queries, workflows or follow-up observations based on new information received. The eSTAR architecture represents a ``turn-key'' system for autonomous observations of transient events, and therefore is ideal for microlensing follow-up.

The agents are also capable of responding in real time to external alerts
\citep{WilliamsSeaman,White},
so-called Virtual Observatory Events (VOEvents)\footnote{{\tt http://www.voevent.org/}}.
While OGLE and MOA alerts are being translated into this format, the detection of an
anomaly by {\sc SIGNALMEN} will also be reported by means of a VOEvent.

Besides the communication by means of software agents, the design of
the {\sc SIGNALMEN} anomaly detector also contains interfaces for output
to human observers and upload of data provided by any other observing campaign. Currently, data from PLANET, OGLE, MOA,
and MicroFUN are fed in. Moreover, we will keep two separate mailing lists for notification on the decision in favour of an ongoing anomaly ('anomaly' status) and on the detection of deviant points ('check' status), which everyone is free to subscribe to. 
While dense follow-up by other teams is much encouraged in this case,
the 'check' status will be invoked frequently (several times each night)
and mainly serves to steer the internal feedback with the robotic telescopes of the RoboNet network and in second instance with the other telescopes involved in the PLANET/RoboNet campaign.
In addition to providing real-time notification of suspected or ongoing
anomalies, we will publish up-to-the-minute plots showing collected
data along with a model light curve, whose parameters have been
determined by the {\sc SIGNALMEN} anomaly detector.

On the suspicion of an anomaly, a fast response with further observations
is crucial for either confirming or rejecting this hypothesis. While 
robotic telescopes can react almost instantaneously, human observers 
need to be informed by e-mail or other means of communication, which adds
some delay. Only if an observatory is staffed and the observer frequently
monitors incoming e-mail, the feedback loop can be closed.
This works reasonably well with the current PLANET network, where  observers are present at the telescope on each night with suitable
weather during the observing season. However, telescopes that are only
activated on a target-of-opportunity basis, such as several of those 
used by MicroFUN, might miss the short-notice call.
In any case, the success of the strategy is limited by the need to 
find out whether a suspected anomaly is present or not with the use
of telescopes that have already monitored the microlensing event of
interest. The value of data from other sites is limited to providing
early useful data if it turns out that an anomaly is ongoing, but these
contain rather little information about whether the light curve deviates.

While so far, we have implemented an algorithm that alerts us on
suspected or ongoing anomalies, it neither gives us a recommendation of the best anomaly sampling interval, for which we simply assume an initial 
choice of 10~min, nor does it inform us when the anomaly is
over and we can return to the standard follow-up sampling rate.
Both of these issues currently need to be dealt with by human interaction
through an internal webpage automatically listing events that
are considered to deviate from ordinary light curves.

\section{The anomaly detector algorithm}
\label{sec:algo}

\subsection{Basics, data statistics, and deviations}

The implementation of the {\sc SIGNALMEN} anomaly detector described in
the following is a first sketch,
matching the primary requirements.
It involves some basic statistical tests, building upon prior experience.
More sophisticated tests can be designed and added, should it turn out that these yield significant improvements on the decision
process.
During the 2007 season, {\sc SIGNALMEN} will log all incoming data, the anomaly indicators, current model parameters, and its decisions, 
which will provide a valuable basis for further tuning.
Our algorithm involves several constants that can be adjusted.
 Their values can be changed by editing a configuration file rather than requiring alteration of the source code itself. In the
following, we list our default setting in brackets.

With the source and background fluxes, $F_\mathrm{S}^{[s]}$ and
$F_\mathrm{B}^{[s]}$, depending on the data archive $s$, residuals 
need to be compared by means of the magnifications
\begin{equation}
A_i = \frac{F_i-F_\mathrm{B}^{[s(i)]}}{F_\mathrm{S}^{[s(i)]}}
\label{eq:Ai}
\end{equation}
rather than the measured fluxes $F_i$, where the uncertainties of
$A_i$ are given by
\begin{equation}
\sigma_{A_i} = \sigma_{F_i}/|F_\mathrm{S}^{[s(i)]}|
\label{eq:sigmaAi}
\end{equation}

In general, the reported error bars $\sigma_{F_i}$ are not a proper
reflection of the true scatter, which moreover frequently deviates
from a Gaussian distribution. In particular, data provided by OGLE come with severely underestimated photometric uncertainties for $I \leq 15$, whereas these are about the right size for $15 \leq I \leq 18$ and overestimates for faint targets $I \geq 18$. One of the sources of this
behaviour is that the photometric reduction packages usually do not take into account further systematic uncertainties. We therefore correct for this fact by adding a systematic error {$\tt SYST\_ERR$} (0.003) in quadrature to the uncertainty of the reported magnitude.
Moreover, rather than relying on $\sigma_{F_i}$, we assess the scatter by means of two statistics, namely the median scatter $\hat \delta^{[s]}$ and the critical scatter $\delta_\mathrm{crit}^{[s]}$. By calculating the 
residuals
\begin{equation}
\delta_k = \frac{A(t)-A_k}{\sigma_{A_k}}
\end{equation}
for each archive $s$ and sorting the $n^{[s]}$ values 
$\left(\delta_k^{[s]}\right)^2$
in ascending order, we 
find
\begin{equation}
\hat \delta^{[s]} = 
\left\{\begin{array}{l}
\left[\left(\delta_{(n^{[s]}+1)/2}^{[s]}\right)^2\right]^{1/2} 
\hfill \mbox{for} \quad
n^{[s]} \;\mbox{odd} \\
\left\{\frac{1}{2}\left[\left(\delta_{n^{[s]}/2}^{[s]}\right)^2
+\left(\delta_{n^{[s]}/2+1}^{[s]}\right)^2\right]\right\}^{1/2} \\
\hfill \mbox{for} \quad
n^{[s]} \;\mbox{even} 
\end{array}
\right.\,,
\end{equation}
and with the critical real index
\begin{equation}
\nu^{[s]} = n^{[s]}\,(1-{\tt DEV\_PERC})+1
\end{equation}
we determine
\begin{equation}
\delta_\mathrm{crit}^{[s]} = \left\{
\begin{array}{l}\left[\left(\delta_{n^{[s]}}\right)^2\right]^{1/2} 
\hfill
\mbox{for} \quad \nu^{[s]} \geq n^{[s]} - 1\\
\left[\beta^{[s]}\,\left(
\delta_{\lceil\nu^{[s]}\rceil}\right)^2 +
(1-\beta^{[s]})\,
\left(\delta_{\lfloor\nu^{[s]}\rfloor}\right)^2 \right]^{1/2} \\
\hfill
\mbox{for} \quad \nu^{[s]} < n^{[s]}-1\\
\end{array}\right.\,,
\end{equation}
where $\beta^{[s]} = \nu^{[s]}-\lfloor\nu^{[s]}\rfloor$,
{$\tt DEV\_PERC$} (0.05) denotes the probability for an absolute deviation
in excess of $\delta_\mathrm{crit}^{[s]}$, and
$n^{[s]}$ is the number of data points for archive $s$.
With a deviation threshold {$\tt DEV\_SIG$} (2), we require for a
significant deviation both
\begin{equation}
|A_j-A(t_j)| > {\tt DEV\_SIG}\, \sigma_{A_j}\,\max\left\{1,\hat \delta^{[s(j)]}\right\}
\label{eq:dev1}
\end{equation}
and
\begin{equation}
|A_j-A(t_j)| > \sigma_{A_j}\,\delta_\mathrm{crit}^{[s(i)]}
\label{eq:dev2}
\end{equation}
to hold. For Gaussian errors bars, both conditions can be made to coincide. 
In order to allow for a proper evaluation of the scatter, we require that
at least {$\tt MIN\_DATA\_TEST$} (6) data points and data from at least 
{$\tt MIN\_NIGHTS$} (2) previous nights have been collected. Otherwise,
the statistical scatter is likely to be underestimated and therefore false alerts are almost certain.

With our robust-fitting algorithm that downweighs or even ignores outlier
and the fact that we rely on the median scatter and trigger on the
absolute residual exceeding that of a fixed percentage of data, we are
well able to distinguish between low-quality data and real deviations.
In particular, this allows us to achieve a low false alert rate.
The requirement of obtaining significant data statistics before assessing
deviations come at the price of some inability to identify deviations in fast-rising events with high-magnification peak. However, this does not significantly affect the planet detection prospects, since
a high-cadence sampling will be carried out for these events 
irrespective of suspected anomalies in the data.

\subsection{Data sequence and modelling}

{\sc SIGNALMEN} assumes that events do not exhibit anomalies at the time
these are first announced by the OGLE or MOA microlensing surveys.
For each data archive we keep track
of the latest collected data point and restart our assessment for
anomalies at the epoch $t_\mathrm{new}$ corresponding to the
earliest 'new' point among all archives. In order to assess the data
point by point, we sort these in time sequence and step 
through points $k \leq n$ with $t_\mathrm{k} \geq t_\mathrm{new}$,
where $n$ is the index of the most recently observed data point.
For each event, we store the time ranges for which anomalies were
considered to be ongoing, and the parts of these ranges prior to
$t_\rmn{new}$ are then excluded from fits for models of an ordinary
light curve. Moreover, on each run of {\sc SIGNALMEN} on a specific
event, we note the epoch $t_\rmn{c} \geq t_\rmn{new}$ for which an ongoing anomaly was first suspected, and administrate a list of all points $l$ 
with $t_\rmn{l} \geq t_\rmn{c}$ that were found to deviate, which form
the current anomaly sequence. When considering all data with $t \leq t_k$,
the deviation of a point with index $j$ ($t_\rmn{c} \leq t_j \leq t_k$)
can be determined with respect to the following models that 
include all data with indices $i$ that fulfill:
\begin{itemize}
\item{{\em 'previous'}\/:} $t_i < t_\mathrm{k}$, exclude data within
	an anomaly time range prior to $t_\rmn{new}$ or in the current  anomaly
        sequence
\item{{\em 'current'}\/:} $t_i \leq t_\mathrm{k}$, exclude data within
	an anomaly time range prior 
        to $t_\rmn{new}$ or in the current anomaly
        sequence 
\item{{\em 'all non-deviant'}\/:} $t_i \leq t_\mathrm{k}$,
        exclude data within
	an anomaly time range prior to $t_\rmn{new}$
    or in the current anomaly
        sequence, but include $i = j$
\item{{\em 'all-data'}\/:} $t_i \leq t_\mathrm{k}$, exclude data within 
        an anomaly time range prior to $t_\rmn{new}$
\end{itemize}
If there is no current anomaly sequence, i.e.\ none of the points $k \leq n$ has been found to deviate so far, the {\em 'all-data'} and {\em 'all non-deviant'} models coincide with the {\em 'current'} model. 
Since model predictions can be expected to fail frequently, our initial
assessment of a deviation is solely based on the {\em 'current'} model,
which includes the latest considered point $k$. Should this point
fail to deviate significantly by means of the conditions given by
Eqs.~(\ref{eq:dev1}) and~(\ref{eq:dev2}), the {\em 'current'} model
becomes the {\em 'previous'} model and $k$ is increased. 
Otherwise, $t_\rmn{c} \equiv t_\rmn{k}$ and data point $k$ is added
to the current anomaly sequence. While the {\em 'previous'} model is
retained, it also becomes the {\em 'all non-deviant'} model, whereas
the {\em 'current model'} also becomes the {\em 'all-data'} model. 
For increased $k$, further tests will be performed for data
$j$ ($t_\rmn{c} \leq t_j \leq t_k$).

\subsection{Anomalies: accept or reject?}

If a current anomaly sequence has been found, {\sc SIGNALMEN} will try
to figure out whether further data points provide evidence in favour
of an ongoing anomaly or against it, leading to finishing up with
'anomaly' or 'ordinary' status. If the current data do not allow to
arrive at either of these conclusions, the 'check' status is invoked.
In this case, the markers for the latest data points for each of the archives are set so that the current anomaly sequence is reassessed
on the next run of {\sc SIGNALMEN}. This avoids the necessity to store
further information about this sequence and also easily allows for a 
potential revision of these critical data in the meantime. 

Data taken after $t_\rmn{c}$ that are found not to deviate significantly
from the {\em 'current'} model can provide evidence against the 
presence of an ongoing anomaly. However, simply counting the number of
non-deviant points is not a viable option since these might have larger uncertainties than the deviant points. This
happens in particular if later data originate from different sites, while even for the same site it cannot be guaranteed that the same 
photometric uncertainty can be retained. Since data with
large scatter and therefore no indication of an anomaly must not be
used as evidence against, it is unavoidable that the
photometric uncertainties are taken into account. Moreover,
we also need some characteristic for the amplitude of the anomaly which
we would like to decide about whether it is real or not. Let us consider
the fractional deviation 
\begin{equation}
\varepsilon_i = \frac{A_i-A(t_i)}{A(t_i)}\,,
\end{equation}
and for a deviant point $l$ define $\varepsilon_l$ as the anomaly level.
With $\sigma_{\varepsilon_j} =  (\sigma_{A_j}\,\max\left\{1,\hat \delta^{[s(j)]}\right\})/A(t_j)$, we then obtain the
weighted average over all non-deviating points $j$ after the deviant
point (i.e.~$t_j > t_l$) 
\begin{equation}
\overline{\varepsilon} = \frac{\sum\frac{\varepsilon_j}
{\sigma_{\varepsilon_j}^2}}{\sum\frac{1}{\sigma_{\varepsilon_j}^2}}\,.
\end{equation}
and its standard deviation
\begin{equation}
\sigma_{\overline{\varepsilon}} =
\left(\sum\frac{1}{\sigma_{\varepsilon_j}^2}\right)^{-1/2}\,.
\end{equation}
The anomaly is then called off if
\begin{eqnarray}
\overline{\varepsilon} & < & \varepsilon_l/2 - {\tt REJECT\_SIG}\;
\sigma_{\overline{\varepsilon}} \qquad
(\mbox{for}\quad\varepsilon_l > 0) \nonumber \\
\overline{\varepsilon} & > & \varepsilon_l/2 + {\tt REJECT\_SIG}\;
\sigma_{\overline{\varepsilon}} \qquad
(\mbox{for}\quad\varepsilon_l < 0)
\end{eqnarray}
with a default setting ${\tt REJECT\_SIG} = 1$ and
the additional requirement that at least {$\tt MINPTS\_REJECT$} (4)
non-deviant points have been collected. For Gaussian data with constant
error bars, we find the anomaly call-off typically not requiring more
than 5 measurements. However, this can take significantly longer if
only data with large effective error bars (corrected for actual scatter)
can be acquired.

If the data point $k$ has been found not to deviate, we also reassess
the current anomaly sequence with respect to the {\em 'all non-deviant'}
model. If an anomaly cannot be confirmed or discarded, just
testing points in sequence against the {\em 'current'} model
can either lead to missed anomalies
or false alerts if the model is not well-constrained. We therefore
determine the residuals with respect to a model that includes all points found not deviating (and their scatter). This also allows taking into account an increased scatter present in more recent data.
Previously deviant data that do not fulfill the new criterion are
removed from the current anomaly sequence, which might lead to a revision
of $t_\rmn{c}$ and leave {\sc SIGNALMEN} with an empty current anomaly
sequence. In the latter case, {\sc SIGNALMEN} will continue as if no deviant points were found in the current run. We also require that all
data points in the current anomaly sequence deviate to the same side. Therefore, it will be shortened if necessary to meet this condition.

Similarly, if the most recently considered data point $k$ is found to deviate to the opposite site as the previous data, a new current anomaly sequence is started at $t_\rmn{c} \equiv t_k$ and the previous sequence is
abandoned.

A stronger hint for an anomaly being ongoing is obtained if the
data point $k$ deviates to the same side as the previous points
in the current anomaly sequence. Once the current anomaly sequence 
contains at least two data points, we start testing the collected
data against an {\em 'all-data' model}, which also contains the
points in the current anomaly sequence. With the earlier
tests we avoided that the model of an ordinary event is driven towards
points that deviate from it, which allows us to  call off an anomaly if further points follow an ordinary light curve without getting confused by outliers. However, we also need to take care of the fact that more weight than just that of a single point might be needed to correct for a bad earlier estimate of model parameters. As a compromise, we adopt
less strict criteria, namely that the residuals of the
last $\tt MINPTS\_ANOMALY$ (5) points are all of the same sign
and at least $\tt MINPTS\_ALL\_ANOM$ (3) points
deviate significantly. If earlier data in the current anomaly 
sequence cannot match these criteria, the sequence is shortened
and $t_\rmn{c}$ is revised.

A further test is based on the overlap between the points in the
current anomaly sequence and non-deviant points falling in between.
With the {\em 'all-data'} model, we determine 
\begin{equation}
\delta A_i = A_i - A(t_i)\,.
\end{equation}
If for a non-deviant point $j$ following a deviant point $l$ for 
which $\delta A_l > 0$, one finds
\begin{equation}
\delta A_j + 2 \sigma_{A_j}\,\max\{1,\hat\delta^{[s(j)]}\}
 < \delta A_l - 2 \sigma_{A_l}\,\max\{1,\hat\delta^{[s(l)]}\}
\end{equation}
or the equivalent relation to hold for the subsequent deviant point,
the non-deviant point is considered to contradict point $l$ deviating,
which is therefore removed from the current anomaly sequence.
For  $\delta A_l < 0$,\footnote{Obviously, there is no $\delta A_l = 0$
case.} the corresponding condition reads
\begin{equation}
\delta A_j - 2 \sigma_{A_j}\,\max\{1,\hat\delta^{[s(j)]}\}
 > \delta A_l + 2 \sigma_{A_l}\,\max\{1,\hat\delta^{[s(l)]}\}\,.
\end{equation}

Finally, we realize that the photometric reduction might fail occasionally
and produce weird results. A common characteristic that can be distinguished
from real anomalous behaviour are sudden changes between a rise and
fall. We therefore determine the pattern of significant increase or
decrease of the magnification amongst the data in the current anomaly
sequence. Should there be more than one change in direction, 
{\sc SIGNALMEN} abstains from the claim that an anomaly is ongoing.
This 'zig-zag test' is only used as the final criterion once all 
other conditions for an ongoing anomaly are fulfilled. 
For two deviant points $l$ and $m > l$, a significant increase
is characterized by
\begin{equation}
\delta A_m - 2 \sigma_{A_m}\,\max\{1,\hat\delta^{[s(m)]}\}
> \delta A_l + 2 \sigma_{A_l}\,\max\{1,\hat\delta^{[s(l)]}\}\,,
\end{equation}
whereas a significant decrease arises by exchanging $l$ and $m$.
If there is no significant change between neighbouring points,
a significant increase is assessed with respect to the lowest of these
points while a significant decrease refers to the highest of these.

To summarize, {\sc SIGNALMEN} concludes that there is an ongoing anomaly
if all of the following criteria are satisfied
\begin{itemize}
\item the anomaly is not called off by means of a series of at least
{$\tt MINPTS\_REJECT$} (4)
non-deviant points with a weighted-average fractional deviation
significantly (measured by $\tt REJECT\_SIG$ (1.0)) closer to zero than half of the fractional deviation of the previous deviant point
\item the most recent deviant points form a sequence of at least
$\tt MINPTS\_ANOMALY$ (5) points
 that were found to deviate to the same side
from the {\em 'current'} model and the {\em 'all non-deviant'} model
\item the residuals with respect to the {\em 'all-data'} model of
at least the last $\tt MINPTS\_ANOMALY$ (5) points in the current anomaly are all of the same sign
\item at least $\tt MINPTS\_ALL\_ANOM$ (3) points in the current anomaly
     sequence deviate from the {\em 'all-data'} model
\item there are no non-deviant data points in between those in the
   current anomaly sequence that significantly fail to overlap with them
\item data in the current anomaly sequence do not change more than
     once between a significant increase and decrease
\end{itemize}

If these criteria are fulfilled for $k = n$, i.e.~at the end of the
collected data, {\sc SIGNALMEN} activates the 'anomaly' mode. Should
these be fulfilled earlier ($k < n$) only, {\sc SIGNALMEN} finishes
with 'ordinary' status, but a file notifying about a missed anomaly
is written. If just the zig-zag test fails, {\sc SIGNALMEN} notifies
about problems with the photometric reduction and suspends evaluation
of data archives for which recent data showed more than one change
of direction in the suspected anomaly sequence. Such a case needs
human intervention and should be dealt with at high priority.

\section{Prospects with the anomaly detector}
\label{sec:prospects}

\begin{figure}
\includegraphics[width=84mm]{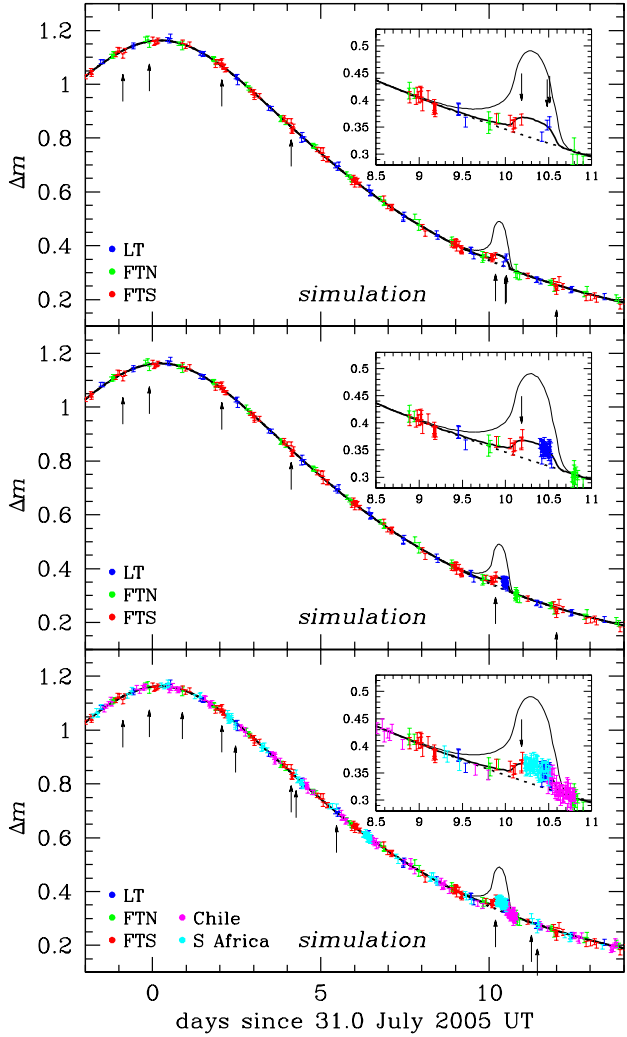}
\caption{Detection of a hypothetical Earth-mass planet in 
microlensing event OGLE~2005-BLG-390 located in the same spot
as OGLE~2005-BLG-390Lb with the RoboNet telescope network. Simulated
data for the different telescopes are shown in different colours along
with theoretical light curves, where a thin solid line corresponds to
the actually derived  model for that event (with OGLE~2005-BLG-390Lb),
a bold solid line to a model with an Earth-mass planet, and a bold
dotted line to a model without planet. Arrows mark data points that
have been found to deviate from the best-fitting model available 
at that epoch. While the top panel shows only data collected with
the standard sampling, the middle and bottom panel include further
data with high-cadence (10 min) sampling after having triggered on
the anomaly. For the top and middle panel, only the current 
RoboNet-1.0 telescopes have been considered, whereas for the bottom 
panel, the availability of two additional similar telescopes in Chile and South Africa has been assumed.}
\label{fig:OB05390simu}
\end{figure}

In order to demonstrate what can be achieved with the anomaly detector,
let us use the event OGLE~2005-BLG-390, which already allowed us to
detect a planet of $5~M_\oplus$ (with a factor two uncertainty), as
an illustrative example and starting point of the discussion.
Fig.~\ref{fig:OB05390simu} shows the model light curve for the
corresponding configuration again, where the planet OGLE~2005-BLG-390Lb
has been replaced by a $1~M_\oplus$ version in the same spot, but rather
than the collected data, we now show 
simulated data related to the three robotic 2m telescopes that currently
comprise the RoboNet-1.0 network: the Liverpool telescope (LT), the
Faulkes telescope North (FTN), and the Faulkes telescope South (FTS).
According to the target observability from the different sites
at the event coordinates 
($\mbox{RA} = 17\fh{}54\fm{}19\fs{}19$, $\mbox{Dec} =
-30\fdg{}22\farcm{}38\farcs{}3$ (J2000)), requiring that
the target is at least $20\degr$ above the horizon, synthetic data
have been generated where the average sampling interval is 
$\Delta t = (2~\mbox{h})/\sqrt{A}$ and the photometric accuracy is 
1 per cent
at baseline and smaller as the event brightens, following photon noise
statistics, where Gaussian errors have been assumed. A systematic
error of 0.5 per cent has been added in quadrature. For the time
of the next observation, a Gaussian fluctuation of 20 per cent of the
sampling interval has been adopted, and while its photometric uncertainty
itself fluctuates by 12.5 per cent. Moreover, a drop-out probability of 5 per cent on each data point has been assumed.

While it would have been rather easy to detect and characterize 
a planet like OGLE 2005-BLG-390Lb, the standard sampling would have
given rather little evidence for an Earth-mass planet in the same
spot, and a characterization would not have been possible.
Arrows in Fig.~\ref{fig:OB05390simu} indicate data points that
were found to deviate by the 
anomaly detector, given the best-fitting model that could have been
obtained at that time, based on all previous data. 
Further data after the first four trigger points would have indicated
that there is no ongoing anomaly, but the deviation is rather of
statistical nature. In contrast, subsequent data points after the
first trigger point that falls onto the real anomaly would have 
confirmed the deviation and finally led to the activation of 
high-cadence anomaly monitoring. This example however also shows
a critical weakness of the current RoboNet-1.0 network, namely its
lack of round-the-clock coverage. In particular, one sees that
the southern telescope offers a much longer time window for our
purpose than either of the northern telescopes. Just after the
opportunity of taking a further point after activation of
the 'check' status, the target could not have been followed anymore
with the FTN and it would have been necessary to wait for the LT for acquiring
subsequent measurements. This demonstrates that provision of
a fast response also implies the availability of a telescope at the 
requested time. Further telescopes available in South Africa and
Chile (see Fig.~\ref{fig:OB05390simu}) would have allowed 
a coverage of the anomaly sufficient to detect an Earth-mass planet.

\begin{figure}
\includegraphics[width=84mm]{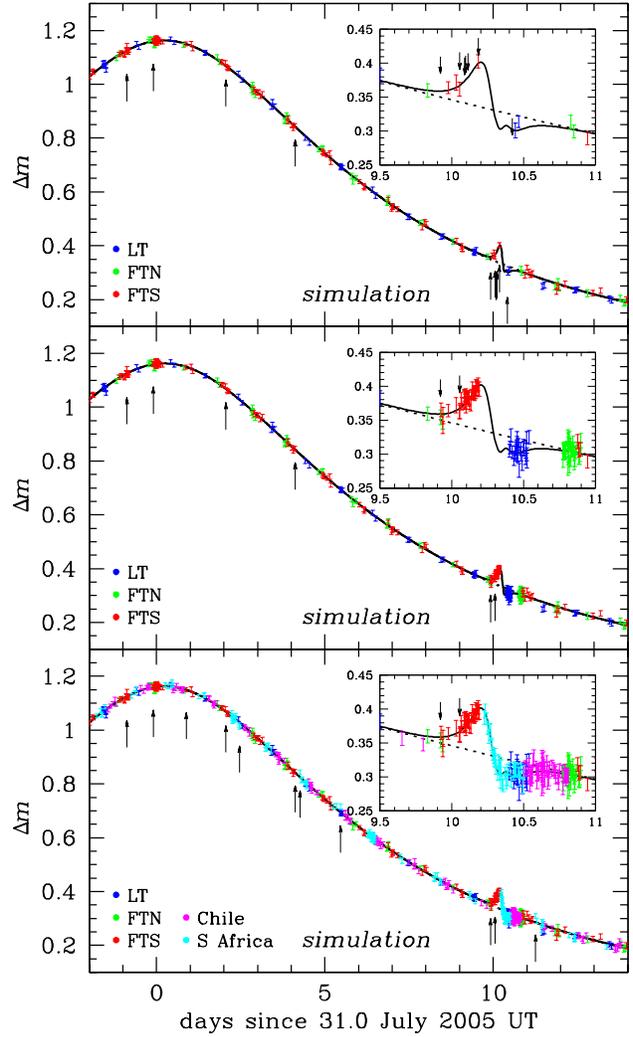}
\caption{Detection of a hypothetical Earth-mass planet in an event
resembling OGLE~2005-BLG-390 at the same separation as OGLE~2005-BLG-390Lb,
but for a main-sequence source star with $R_\star \sim 1.2~R_{\sun}$.
In fact, the source size has been assumed to be 8 times smaller than for the giant observed in OGLE~2005-BLG-390. Since with the smaller source, the planet is missed for the original angle $\alpha = 157.9\degr$ 
\citep{PLANET:planet} between the line from the planet to its host star and the source trajectory, where the lens centre of mass is to the right, a slightly different angle $\alpha =  158.2\degr$ has been adopted, for which a 5 per cent deviation results. Otherwise, this figure is analogous
to Fig.~\ref{fig:OB05390simu}.}
\label{fig:OB05390simuMS}
\end{figure}

\begin{figure}
\includegraphics[width=84mm]{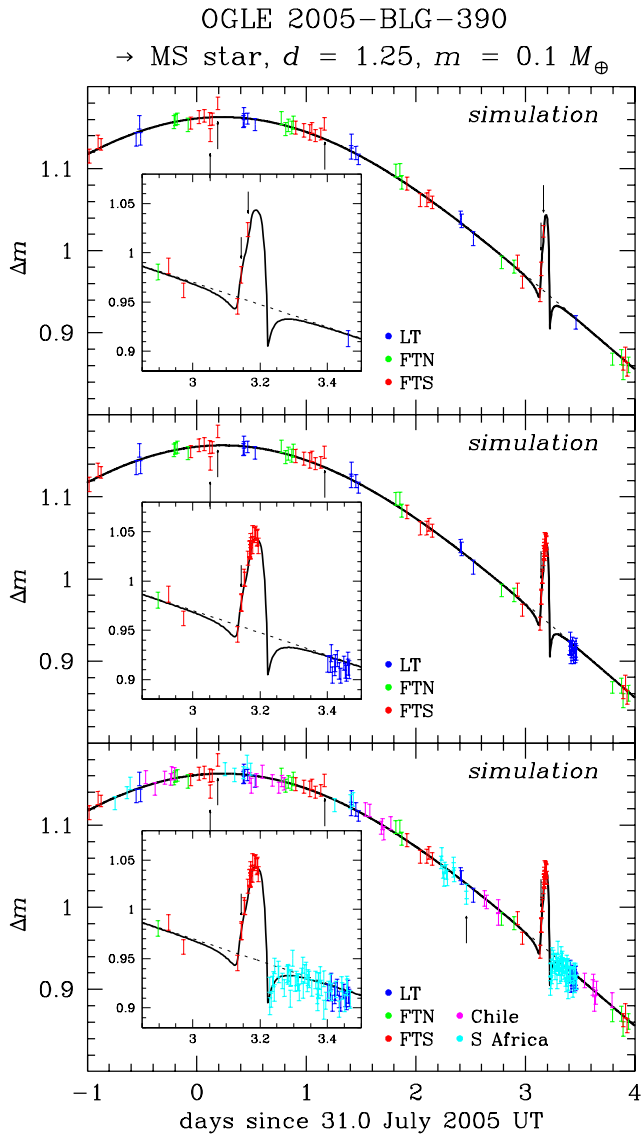}
\caption{Coverage of a microlensing event resembling OGLE~2005-BLG-390
with a  main-sequence source star ($R_\star \sim 1.2~R_{\sun}$)
instead of the 8 times larger giant, and a planet of $0.1~M_{\oplus}$ at
an angular spearation of $d = 1.25$ times the angular Einstein radius
$\theta_\rmn{E}$. The angle between planet-to-star axis and source 
trajectory, with the lens star to the right, has been chosen 
as $\alpha = 126.6\degr$, so that the source crosses the planetary
caustic. As for Fig.~\ref{fig:OB05390simu}, the trigger points of the
anomaly detector are indicated by arrows, and the middle and bottom
panels include the high-cadence follow-up after anomaly suspicion or
detection. While the top and middle panels only show data corresponding
to the current RoboNet-1.0 network, the availability of further similar
telescopes in Chile and South Africa has been assumed.}
\label{fig:OB05390subsimu}
\end{figure}

The detection of OGLE~2005-BLG-390Lb was eased by the source
being a giant star with a radius $R_\star \sim 9.6~R_{\sun}$, which not
only allowed to obtain an accurate photometry with rather short exposures
but also increased the probability of detecting a significant deviation.
While the large angular size led to a reduction of the amplitude, which
did not matter because it remained at the 15 to 20 per cent level,
a larger range of orientations or impact parameters than
for a smaller source would have created on observable signal. Moreover,
the duration of the planetary deviation is dominated by the source star
moving by its angular size, giving a rather comfortable timespan, which
would still have been $\sim 12~\mbox{h}$ for an Earth-mass planet.
While a main-sequence star could have provided a signal with larger 
amplitude, the probability to observe it would have been smaller and
it would not have lasted for that long.

If one replaces the source star of OGLE~2005-BLG-390 with an 8 times
smaller version, an Earth-mass planet in the same spot as
OGLE~2005-BLG-390Lb would become undetectable since the smaller source would not enter a region for which significant deviations result.
For the configuration shown in Fig.~\ref{fig:OB05390simuMS}, the 
angle of the source trajectory relative to the planet-star axis has 
therefore been slightly adjusted, resulting in a 5 per cent deviation.
Achieving good photometry on the fainter target is more difficult and
requires longer exposure times. Nevertheless, PLANET has demonstrated
a photometric accuracy of even less than 0.5 per cent on a main-sequence star is possible provided that it is fairly isolated
rather than in a crowded area.
While for the previously discussed case involving a giant source star, signal amplitudes significantly exceeding 3 per cent cannot result, 
the shown 5 per cent deviation is not even near the limit for 
main-sequence stars, for which very strong signatures become possible 
should the source happen to cross a caustic.
One also sees that the duration of the planetary deviation has not
decreased by a factor of 8 as the source size did. Contrary to the
giant source star case, the signal duration is now roughly given by
the time in which a point source passes the region of angular positions
that lead to significant deviations, and remains $\sim 8~\mbox{h}$ for
the prominent peak. The angular size of the 
source star itself is reflected in the small peak within the timespan
over which the brightness in presence of the planet is smaller than without. As before with a giant source star, the proper characterization
of the planetary anomaly is not possible with the standard 
2~\mbox{h} sampling, while high-cadence sampling after having suspected
or detected an anomaly will solve the problem, provided that telescopes
are available to observe the target. Interestingly, in a very early
stage of the anomaly, one of the data points appears to be higher just
by chance. Further data taken at the sampling interval of 10~min however
do not confirm a significant deviation, so that only after the
next data point taken with the standard sampling rate the high-cadence
anomaly monitoring remains active.

After having found that the discovery of Earth-mass planets 
does not constitute the limit of what can be achieved with microlensing
survey/follow-up campaigns equipped with an automated anomaly detector,
let us look into how far one can go. In fact, the rather large
separation $d \sim 1.6$ of OGLE~2005-BLG-390Lb from its host star
did not offer a very fortunate configuration.
Let us therefore also consider $d \sim 1.25$ and see how the signal amplitude and duration are affected.
As Fig.~\ref{fig:OB05390subsimu} shows, even for a planet with mass
$m = 0.1~M_{\oplus}$ located at $1.25\,\theta_\rmn{E}$ from its host
in an OGLE~2005-BLG-390-like event, a signal of 10 per cent lasting 
about 2.5~hrs can result on a main-sequence source star. 
The early detection of such a short signal with a standard survey
sampling interval of 1--2~hrs and the anomaly monitoring each become challenging. In fact, the rather short time gap of about 40~min between
the FTN in Australia and a telescope in South Africa is sufficient for
missing the crucial falling part of the planetary anomaly in our
simulation. This also demonstrates the extraordinary value of a
telescope at the western edge of Australia and/or in southern central Asia.
Nevertheless, the discovery of planets with masses of even
$0.1~M_{\oplus}$ or below by ground-based microlensing campaigns
remains a matter of probability rather than possibility.

\section{Summary and conclusions}
\label{sec:summary}
For probing models of planet formation and evolution and thereby
taking an important step towards our understanding of the
origin of living organisms as we know them, microlensing will remain
a competitive and complementary technique with respect to other
methods for the foreseeable future
with its unique sensitivity to low-mass planets in wider orbits.
It is unlikely (although not impossible) that microlensing will provide a timely discovery of a planet on which conditions similar to those on Earth can exist that are known to support the formation of life forms,
with less than 3 per cent of planets in any mass range expected to orbit
suitable host stars at suitable radii \citep{Han:habitable}.
While both transit and radial-velocity surveys approach the
required orbital range for such habitable planets from closer
orbits, essentially all planets that can be expected to be detected by microlensing reside in wider orbits. The discovery of the first
extra-solar planets already demonstrated impressively how little one can infer about the origin of the Solar system if the study remains
restricted to itself.
Similarly, one should not expect that a study just of habitable planets
will allow to arrive at a well-understood
picture of their formation. Instead, a reliable test of theories should 
involve data spanning over a
wider and surrounding region. Moreover, microlensing
allows to obtain a planet sample not only around stars in the Galactic
disk but also around those the Galactic bulge, thereby probing two
distinct populations.

We have shown that our {\sc SIGNALMEN} 
anomaly detector, which went into operation  
for the 2007 microlensing observing season, allows to adopt a
3-step strategy of survey, follow-up, and anomaly monitoring.
The basis of this strategy is formed by the microlensing events
provided in real-time by the OGLE and MOA surveys out of which
a sufficient number are then monitored with a network of 2m telescopes so that deviations due to Earth-mass
planets are unlikely to be missed and a reasonable number of low-mass
planets is expected be detected over the next few years. In particular, it
is only required that the follow-up observations provide an early enough
trigger of an ongoing anomaly, whereas a proper characterization need not
be ensured by these, because this will be achieved by the high-cadence
anomaly monitoring after an anomaly has been suspected or detected.
In fact, the use of an anomaly detector becomes quite efficient if
many points are required for proper characterization of a signal
rather than being able to claim a detection from a single deviant point.
The expected detections will provide a powerful test of models of planet formation and evolution around K- and M-dwarfs.
While planets of Earth mass appear to have some particular appeal,
they do not provide the hard limit for ground-based microlensing 
searches. As shown by one of the examples discussed in Sect.~\ref{sec:prospects}, the anomaly detector allows us to go even
below $0.1~M_{\earth}$, although such detections are not very likely
to occur. However, these are reasonable goals for 
a network of ground-based wide-field telescopes or a space-based telescope
\citep{Bennett:space}. 

By only checking for significant deviations from an ordinary microlensing
light curve, our current anomaly detector is blind to the nature of origin
of the observed deviation. While it is ensured that planetary deviations
due to planets of Earth mass and even below can be detected, more than
90 per cent of all deviations are due to other causes,
such as finite-source effects, Earth-sun parallax,
or stellar lens or source binaries. 
In order to distinguish these from planetary deviations and at the
same time to obtain appropriate estimates for the urgency and frequency of second-level follow-up observations, a full real-time modelling taking into account all these effects would be required. Optimally, the prioritization
of events would follow the expected constraints on the planet characteristics rather than just maximizing their detectability while
ignoring the chances of properly characterizing a potential planet.
We plan to implement such a system in the near future.

\section*{Acknowledgments}
We would like to thank K. Horne, M.~F. Bode, S.~N. Fraser, C.~J. Mottram, T. Naylor, C. Snodgrass,
I.~A. Steele, P. Wheatley, J.-P. Beaulieu, D. Bennett, P. Fouqu\'{e}, N. Kains, C. Vinter
and A. Williams for valuable suggestions at various stages.
More thanks go to D. Bennett for providing data files and raw scripts for plotting some of the figures.
We are grateful to the OGLE, MOA, PLANET/RoboNet and MicroFUN teams
for providing real-time photometry on ongoing microlensing events and
communicating recent developments on suspected anomalies. NJR acknowledges financial support by a PPARC PDRA fellowship. NJR and LW were supported by the European Community's Sixth Framework Marie Curie Research Training Network Programme "ANGLES"
(MRTN-CT-2004-505183). EK is supported by a PPARC/STFC Advanced
Fellowship.

\begin{appendix}
\section{The inability to detect planets at the resonant separation
in some events}
\label{sec:appendix}
In general, angular positions for the source star relative to the 
lens composed of a star and its planet, for which a significant deviation
in the observed light curve as compared to a lens star without planet
results, form regions around the caustics of the lens system. 
For point sources, these regions grow as the
angular separation parameter $d$, where $d\,\theta_\rmn{E}$ gives the
angular separation of the planet from its host star, approaches unity.
This indeed implies, as \citet{MP91} pointed out, that planets around
$d \sim 1$ are most easily detected among all events that occur.
However, this does not imply that this also holds for each of the
observed events.

Finite-source effects may cause an increase of the detection efficiency by means of the larger source catching more easily a region of significant deviation without bringing the signal amplitude below the detection threshold, or these can lead to a reduction, in particular if the finite
source subtends regions corresponding to deviations of opposite signs.
In fact, \citet{GauSack:deteff} found that a 
reduction of the detection efficiency most prominently affects separations around the resonant $d \sim 1$, where the wide-separation side suffers more than the close-separation side.
 
In any case, there is already another effect that prohibits the 
detection of less massive planets around $d \sim 1$.
Given that the central caustic is found at the position of the lens
star and the centre of the planetary caustics is separated
by $|d - 1/d|$ from it, the regions for which a significant
deviation results might fall inside a circle whose radius is given
by the impact parameter $u_0$ of a given event. In this case, these
cannot be traced by the source trajectory. For small planet-to-star
mass ratios $q$, the critical separations approach
\begin{eqnarray}
d_{\pm} & = & \frac{1}{2}\left(\sqrt{u_0^2+4} \pm u_0\right) \\
      & \simeq &1 \pm \frac{u_0}{2} \quad (u_0 \ll 1)\,,
\end{eqnarray}
while for larger $q$, an increasingly larger size of the caustics and
the associated regions of significant deviations allows to enter
the range $d \in (d_-,d_+)$ further and further. The fact
that for larger lens-source separations, less massive planets cannot be 
detected in increasingly broader regions around the angular Einstein radius
$\theta_\rmn{E}$ is apparent in the examples showing the detection efficiency as function of the separation parameter $d$ for a few choices of
the planet-to-star mass ratio in Figs.~5 and~8 provided by
\citet{GauSack:deteff}, but unfortunately not discussed there.

\begin{table}
\caption{
Forbidden regions for the planet separation as function
of the event impact parameter}
\label{tab:dcrit}
\begin{tabular}{@{}ccc}
\hline 
$u_0$ & $d_-$ & $d_+$ \\
\hline
1.5 &  0.5  &   2 \\
1.0 &  0.62 & 1.62 \\
0.7 & 0.71  & 1.41 \\
0.5 & 0.78 & 1.28 \\
0.4 & 0.82 & 1.22 \\
0.3 & 0.86 & 1.16 \\
0.2 & 0.90 & 1.10 \\
0.1 & 0.95 & 1.05 \\
\hline
\end{tabular}

\medskip
$d_{\pm}$ are the critical separations 
for which the centre of the planetary caustic(s) falls inside a circle of radius $u_0$ around the host star.
For sufficiently small mass ratios, this will prevent planets 
with separation parameters in the range $(d_{-},d_{+})$ from being
detected in events with impact parameter $u_0$.
\end{table}

Table~\ref{tab:dcrit} shows the critical separations for selected
values of $u_0$. In particular, (low-mass) planets within the lensing zone can only be detected in events with $u_0 \leq 1$ (corresponding to $A_0 \geq 1.34$). Given these findings and the fact that a planet actually did
reveal its existence in the
event OGLE~2005-BLG-390, it is less surprising that with $u_0 = 0.359$,
its angular separation from its host star is the rather large 
$d = 1.61$ angular Einstein radii, whereas a detection at e.g.~$d = 1.1$
would have been impossible.

\end{appendix}

\bibliographystyle{mn2e}
\bibliography{detector}

\end{document}